\documentclass[%
 reprint,
 amsmath,amssymb,
 aps,
]{revtex4-1}

\usepackage{graphicx}
\usepackage{dcolumn}
\usepackage{bm}
\usepackage{natbib}
\usepackage[titletoc,toc,title]{appendix}

\begin{document}

\preprint{APS/123-QED}

\title{Precise Time Evolution of Superconductive Phase Qubits }

\author{Ali Izadi Rad}
 \altaffiliation[Also at ]{Physics Department, Sharif  University Of Technology.}
\author{Hesam Zandi}%
 \email{Zandi@ee.sharif.edu}
 \author{Mehdi Fardmanesh}
 \email{Fardmanesh@ee.sharif.edu}
\affiliation{%
 School of Electrical Engineering, Sharif University of Technology, Tehran, Iran
}%

\collaboration{Superconductor Electronics Research Laboratory (SERL)}

%
%

\begin{abstract}
New procedure on precise analysis of  superconducting phase qubits using the concept of Feynman path integral in qunatum mechanics and quantum field theory has been introduced. 
 the wave-function and  imaginary part of the energy of the pseudo-ground state of the Hamiltonian in Phase Qubits has been obtained from semi-classical approximation and we estimate  decay rate, and thus the life time of meta-stable states using the approach of  Instantons model.  We devote the main  efforts to study the evolution of spectrum of Hamiltonian in time after addition of  interaction Hamiltonian ,  in order to obtain the  high fidelity quantum gates. 

\end{abstract}

\pacs{Valid PACS appear here}
\maketitle


\section{Introduction}
The potential to manipulate information efficiently with
quantum mechanics and remarkable promise of quantum computation  has led to a search and invention  of  a significant number of proposal  for physical system
that could implement a quantum computer in 
large size \citep{107,108,109,110}. Superconducting circuits using Josephson junctions provide
a promising approach towards the construction of a
scalable solid-state quantum computer.These
devices show the quantum effects in macroscopic scale
and it's the most advantage of these elements \citep{106},\citep{111}. They can play the basic building blocks of qunatum computers, which are quibits. Inadition by manupolating the the qubits via external contolled current sources there is possibility to construct the specific quantum gates \citep{112}.

A viable quantum computer needs to has a  the stable and long-live Qubits that make their coherency for long time before the manipulation and operation acts on them.

 Thus
In order to building the quantum bits and quantum gates with high accuracy and high fidelity we need to have a deep recognition to the exact description physics of the system which in translation to quantum mechanics, it mean we should have a precise analysis on evolution of Propagator of the system which is completely time-dependent. 

For the basics operation in qunatum computation we foundamentaly need to study the evolution of N two-level qunatum systems which we can describe their states with the N-component vector $|\psi \rangle $ . The evolution of this state can be given by propagator. In the general, the Hamiltonian is uncomutatable during the time, $[H(t),H(t_0)] \neq 0$ and we are  bound to 
use the approximating method such as Dyson series for finding the  genral form propagator
 Unfortunately it's not easy to calculate epecailly when we need  analytical description of system. In the other hand There are various method to solve the time-independent quantum mechanics problem such as perturbation theory or WKB theory. But in fact in fact WKB is uncontrolled approximation in general and it is hard to say that the result of this methods is accurate or not. Therefor find the methods that help us to get the more accurate and reliable result is very important and essential  

In this paper we claim that functional formalism of
quantum mechanics and Feynman path integral \citep{201},\citep{202} give us
the more accurate answer  about estimating the ground states of energy and describing the meta stable states  wave functions and decay rates of states in superconducting phase qubits. Admittedly the
formalism of path integral has been built completely time
dependent and evolutationary proscess of the system will be tracked more convineint. This is  the biggest treasure that lies down under 
this formalism.

At first section we review on structure of phase potential and we use  from the Instanon model for finding the most properties of ground states of energy up to accuracy of $O(\hbar)$, after that we present the time-dependent propagator of the quantume system and this achevement leads us to find the repersentation of the hamiltonain across the time whie the application of external manipultaion, for the application of building the quantum gates.

\section{Superconducting Phase Qubit}
Single Josephson junctions phase Qubits consists of one Josephson junction which use the quantum tunneling effect to produce the continuous current in the existence of external current source, $I_e$.

 The Hamiltonian of system can be written as 
\begin{equation}
H_{dc}=-E_C \frac{{\partial}^2}{\partial {\delta}^2}+E_J\cos{\delta}+ \frac{\hbar}{2e}I_e \delta
\end{equation}
Where $\delta$  present the phase of Josephson junction and $E_J=\hbar/2eI_e$. In order to manipulate the system we need to evolve the system by time-dependent current, this manipulation introduce the Hamiltonian of interaction which can given by 
\begin{equation}
H_{\mu\nu}=\frac{{\Phi}_0}{2\pi} I_{\mu\nu} \delta=\frac{{\Phi}_0}{2\pi} I(t) \cos(\omega t +\phi ) \delta
\end{equation}
Here ${\Phi}_0=\frac{\hbar}{2e}$ is qunat of flux,Thus the total Hamiltonian of system yields 
\begin{equation}
H(t)=H_0+V(t)=H_0+ \frac{I_0 {\Phi}_0 }{2\pi} I(t) \cos(\omega t +\phi)
\end{equation}


\section{Instanton Model}
In this section we  study the metastable states in tilted-washboard potemtial of Josephson-junction phase qubit. We use from path integral approach for our study. If we consider the particle with unite mass which is under the influence of the one-dimentioanl potential V(x), then following the Euclidean form of the Feynman path integral we can describe the evolution of the particle with 

\begin{equation}\label{9003}
\langle x_f | e^{-\frac{HT}{\hbar}}|x_i \rangle =N \int [dx] e^{-\frac{S}{\hbar}}
\end{equation} 

Here $|x_f \rangle $ and $|x_i\rangle $ are the eigenvalue of the space and the $N$ refer to normalization factor, $H$ represent the Hamiltonian of the system which can be depends on time and $T$ shows the time interval of the evolution.

the symbol of $[x]$ denotes the integration over the all functions $x(t)$ that obey from the boundarycondition $x(-\frac{T}{2})=x_i, x(+\frac{T}{2})=x_f$ If $\bar{x}$ be any functions which obeys the boundary condition then we can write $x(t)=\bar{x}(t) +\sum_{n} c_n x_n(t)$ where the set $x_n$  build the complite set, $\int_{-\frac{T}{2}}^{\frac{T}{2}} x_n(t) x_m(t)={\delta}_{mn}$ and $x_n(\pm \frac{T}{2})=0$. By these condition we can rewrite the mesure of the integral by 
%
%
%
%
%
%

%

Calculation in order of $\hbar$ and using the semi classical approximation lets us to write the evolution of the systme by 
\begin{eqnarray}
\langle x_f | e^{-\frac{HT}{\hbar}} |x_i \rangle &&=N e^{-\frac{S(\bar{x})}{\hbar}} {\prod}_{n} {{\lambda}_n}^{-\frac{1}{2}} [1+O(\hbar)] \\ \nonumber
&&= N e^{-\frac{S(\bar{x})}{\hbar}}[det(-{\partial}_{t}^2+V''(\bar{x})]^{-\frac{1}{2}}[1+O(\hbar)]
\end{eqnarray}

%
%
%
  
 If we define ${\omega}^2$ to be $V''(0)$ then the standard calculation shows that for large $T$ 
 \begin{equation}
N[det(-{\partial}_t^2+{\omega}^2)]^{-\frac{1}{2}}=(\frac{\omega}{\pi \hbar})^{\frac{1}{2}}e^{-\frac{\omega T}{2}}
\end{equation}

\begin{center}
\begin{figure}[h]\label{YaNabi001}
  \includegraphics[height=45mm]{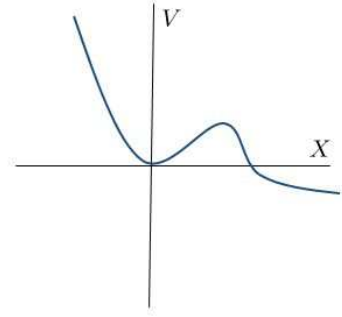}
  \caption{
Special case of potential which provides the bounce motion, it has local minimum, but not the absolute one. 
 }
  \end{figure}
  \end{center}

Now if we consider the desire potential shape as Fig.\ref{YaNabi001}

 Here we still asumme that $x_i=x_f=0$. 
 If we look at the inversed potential($-V(x)$) we can see despite the previous case, here we can have the nontrivial answer.
  the particle can roll itself down from the top of the hill of potential at $x=0$  and goes untill reaches to the turning point. as we interested to limit the time to infinty in the final calculation, then we hold the answers that acuure in long time.  this situation is so called ``the Bounce''. In this situation  Energy of particle is zero. Action in this situation give by 

 \begin{equation}
B=\int_{-\infty}^{\infty}dt (\frac{dx}{dt})^2=\int_{0}^{\sigma} dx [2V(x)]^{\frac{1}{2}}
\end{equation}
where $x=\sigma$ is a place where poential iz zero,$V=0$. If we define the center of the bounce by the place where we have $\frac{dx}{dt}=0$  there, we can see this point is invarinat under time translation. For large time, $T$ , we can put $n$ seprated points like that that have enough space from here and each of points plays the role of one signle Bounce motion. If we show the center of this points by $\frac{T}{2}>t_1>t_2 \cdots > t_n >-\frac{T}{2}$
%
In path integral we should notice that the value of action chanches from $S$ to $nS$ and as this $n$ points are located in the far distance from each other we can write the determinant as the product of determiniat of  many single  Bounce motion. in this way we obtain 
\begin{equation}
(\frac{\omega}{\pi \hbar})^{\frac{1}{2}}e^{-\omega \frac{T}{2}} K^n 
\end{equation}

Where $K$ is a factor that we will discuss on it later. Also the integration over time yeilds 

\begin{equation}
\int_{-\frac{T}{2}}^{\frac{T}{2}} dt \int_{-\frac{T}{2}}^{t_1} dt_2 \cdots  \int_{-\frac{T}{2}}^{t_{n-1}} dt_n =\frac{T^n}{n!}
\end{equation}

finaly we will have 

\begin{equation}
\sum_{n=0}^{n=\infty} (\frac{\omega}{\pi \hbar})^{\frac{1}{2}} e^{-\frac{\omega T}{2}} \frac{(Ke^{-\frac{B}{\hbar}}T)^n}{n!}=(\frac{\omega}{\pi \hbar})^{\frac{1}{2}} e^{[-\frac{\omega T}{2}+Ke^{-\frac{B}{\hbar}}T]}
\end{equation}
 
From Eq.\ref{9003} we can see for example that the correspond value of ground state of enregy can be given by 
\begin{equation}
E_0=(\hbar \omega -\hbar K e^{-\frac{B}{\hbar}})[1+O(\hbar)]
\end{equation}

 We can see that  one of  the eigenvalue is  zero and the corresponding eigenfunction that provide this eigenvalue can be easily given by  $x_1=B^{-\frac{1}{2}}\frac{d\bar{x}}{dt}$,
therefore we need to omitte this trubling zero eigenvlue. we can use many ways to solve this problem and by the standardr way we can calculate prime determinant which the zero eigenvalue has been omitted and we need to add coeficeint to the $K$   like as $(\frac{B}{2\pi \hbar})^{\frac{1}{2}}$ \citep{1009},\citep{1010}.


Now if we review on our solution way with more details we will find that as in one place the $\frac{d \bar{x}}{dt}$   become zero therfor $x$ has   node and thus it cannot be the lowest eigenvalue of enrergy. It means that this system has the negatibe eignevalue and the sepectum of our systems is the special   case and we have unstable state here and the unitarity of this location form the Hilbert sapce can be in doubt.

The key to point for solving  this problem is that  we need to add the coeficeint one-halph to our calucation and the relible result yields \citep{1009},\citep{1010},\citep{1}

\begin{equation}
Im[N \int [dx]e^{-\frac{S}{\hbar}}]=\frac{1}{2}N e^{-\frac{B}{\hbar}}(\frac{B}{2\pi \hbar})^{\frac{1}{2}} T |\text{det}^{'} [-{\partial}_{t}^2+V''(\bar{x})]|^{-\frac{1}{2}}
\end{equation}

and by comparing the result with the defination of $K$ we will find that 

\begin{equation}
Im K =\frac{1}{2} (\frac{B}{2\pi \hbar})^{\frac{1}{2}} |\frac{det'[-{{\partial}_t}^2+V''(\bar{x})]}{det[-{{\partial}_t}^2+{\omega}^2]}|^{-\frac{1}{2}}
\end{equation}

In the stable situation, when the height of barrier penetration goes to infinity the solution of Schrödinger equation, corresponding to the ground state energy $E_0$ behaves as 
\begin{equation}
{\psi}_0(t) \sim e^{-\frac{iE_0t}{\hbar}}
\end{equation}
But for the case that we have not absolute minimum, $E_0$ becomes imaginary. Therefore for long times we have 
\begin{equation}
|{\psi}_0(t)|\sim e^{-\frac{Im E_0 t}{\hbar}}
\end{equation}
It clearly shows that the amplitude and therefore the probability of state decays. The parameter $|\frac{\hbar}{\text{Im}E_0}|$ is the lifetime of a now metastable state with wave function $\psi(t)$ . Let us to point out that the decay of state receives contributions from the continuation of all excited states. However, one expects, for intuitive reasons, that when the real part of the energy increases the corresponding contribution decreased faster with time, a property that can, indeed, be verified in examples. Thus, for large times, only the component corresponding to the pseudo-ground state survives. by considering the one-halp calculation we have 

\begin{eqnarray}\label{9004}
\Gamma && =-2 \text{Im}E_0/\hbar \\
&&=(\frac{B}{2\pi \hbar})^{\frac{1}{2}}e^{-\frac{B}{\hbar}} |\frac{det'[-{{\partial}_t}^2+V''(\bar{x})]}{det[-{{\partial}_t}^2+{\omega}^2]}|^{-\frac{1}{2}}[1+O(\hbar)] \nonumber
\end{eqnarray}

\section{estimation of the coefinceint}

Here we have similar situation to unstable states and bounces, therefore we follow the mentioned solving way that we discussed in previous sections

In order to finding the classical path we should inverse the potential. If we call the turning point by $\sigma$, as is clear in fig 12 , then $\sigma$ is the zero of $V(x)=\alpha\cos{x}+\beta x+\epsilon(x)$ . The analytical solution of this equation obviously is not clear at first sight, specially the correction error, $\epsilon(x)$, has no simple formula. Thus it's better to solve it with soft wares, depend on our parameter. By knowing the turning point then estimating the action, $S_0$ is easy. as usual 
\begin{equation}
S_0=\int_{{\delta}_i}^{{\delta}_f} d\delta\sqrt{\frac{2V(\delta)}{m}}
\end{equation}
From our Hamiltonian it is celear that $m=\frac{{\hbar}^2}{2E_C}$ and $V(\delta)=E_J\cos{\delta}+\frac{\hbar}{2e}I_e \delta+\epsilon(\delta)$. Thus
\begin{equation}
S_0=\int_{0}^{\sigma} d\delta \sqrt{\frac{4E_C}{{\hbar}^2}}\sqrt{E_j\cos{\delta}+\frac{\hbar}{2e}I_e \delta + \epsilon(\delta)+c_0}
\end{equation}
Where $c_0$ is constant that appear form changing the coordinate in order to the hill point of potential locate at zero point of coordinate. Value of this integral can be calculate easily by soft wares.
Now we try to find the classical path. For simplicity and consistency we previous formula in previous subsection we change the variable of motion $x$ or $q$ instead of $\delta$ . From equation of motion we have 
\begin{equation}
\frac{1}{2}m{\dot{x_c}}^2=V(x_c)+E_0
\end{equation}
Thus the classical path obey from this relation
\begin{eqnarray}
t&&=t_1+\sqrt{\frac{2}{m}}\int_{0}^{x'} dx_c \sqrt{V(x_c)+E_0}\\
&&=\int_{0}^{x'} dx_c \frac{\sqrt{\frac{{\hbar}^2}{4E_C}}}{\sqrt{E_J \cos{x_c}+\frac{\hbar}{2e}I_e x_c +\epsilon(x_c)+E_0}}\nonumber 
\end{eqnarray}

The $E_0$ is the constant of motion in must be selected which in $x\rightarrow 0$, $t\rightarrow -\infty$ and vice-versa.
As we know the zero Eigenfunction of $[-{\partial}_t^2+V''(x_c)]$ is 
\begin{equation}
x_1={S_0}^{-\frac{1}{2}}\frac{dx_c}{dt}
\end{equation}
For the next estimation we strongly to know the behavior of $x_1$ respect to time. from previous equation we have function $t(x_c)$ and what we need is the inverse of this function $ g=f^{-1}=x_c(t)$. Finding the analytical form for this function is complicated and it's better to solve it numericaly in exact case that we need and with desire parameters.

%
And for example, for the potentional $V(x)=\frac{1}{2}x^2+\frac{1}{2}g x^4$ the $x_c(t)$ has the form $x_c(t)=g(t) \sim \frac{1}{\cosh(t-t_0)}$, Fig 13.
Hence we expect that $x_1$ behave exponentially when time goes to infinity.
\begin{equation}
x_1=S_{0}^{-\frac{1}{2}}\frac{d x_c}{dt} \rightarrow Ae^{-|t|}, t\rightarrow \pm \infty
\end{equation}
We consider estimating the quantities $S_0$ and $x_c(t)$ let us to estimate the $A$ factor which is constant and fundamentally is function of just $I_C$ and $I_e$ and capacitance and cross section area of Josephson junction.

It is easy to show that in genral \citep{1}
\begin{equation}
\frac{det[-{\partial}_{t}^2+U''(x_c)]}{det[-{\partial}_{t}^2+{\omega}^2]}=\frac{1}{2A^2}
\end{equation}
And 
\begin{equation}
K={(\frac{S_0}{2\pi \hbar})^{\frac{1}{2}}\sqrt{\frac{1}{2A^2}}}
\end{equation}
Thus finally we will have 
\begin{equation}
\Gamma =\hbar (\frac{S_0}{2\pi \hbar})^{\frac{1}{2}}\sqrt{\frac{1}{2A^2}}e^{-\frac{S_0}{\hbar}}
\end{equation}


\section{Properties of Propagator }

Propagator palys the fundematal rules for undersatnding the evolution of the system and the whole properties of the system basicly can obtain from the Propagator and here we want find the element of the Hamiltonian of our sytem which are dependent on time. As we know 

\begin{equation}
\langle x_f |U(t_f,t_i)|x_i \rangle =U(x_f,t_f;x_i,t_i)=\int_{x(t_f)}^{x(t_i)}D[x(t)]e^{\frac{i}{\hbar}S[x(t)]}
\end{equation}

It is clear that the Path Integral are invariant under the transformation  of $x(t) \rightarrow x(t)+y(t) ,y(t_i)=y(t_f)$Which yields to Equation which is so called Schwinger-Dyson equation:

\begin{equation}
\int_{x(t_i)=x_i}^{x(t_f)=x_f} [Dx(t)]\frac{\delta}{\delta x(t) }e^{\frac{i}{\hbar} S[x(t)]}=0
\end{equation}

Which is equal to 
In the other hand as we know 
\begin{equation}
S[x(t)]=\int_{t_i}^{t_f} L(x(t),\dot{x}(t);t)dt
\end{equation}


combining this equation with Schwinger-Dyson equation we found that the Identity 

\begin{equation}
\int_{x(t_i)=x_i}^{x(t_f)=x_f} [Dx(t)]\Big(\frac{\partial L}{\partial x}-\frac{d}{dt}\frac{\partial L}{\partial \dot{x}(t)}\Big) e^{\frac{i}{\hbar} S[x(t)]}=0
\end{equation}

This Identity is some aspect of the Ehrenfest Thoerem.

Now  if looking for the variation od Action with conditions which  $\delta x(t_i)=0 ,\delta x(t_f) \neq 0$
,then we have 
\begin{eqnarray}
\delta S[x(t)]& =\int_{t_i}^{t_f} dt \Big( \frac{\partial L}{\partial x}-\frac{d}{dt}\frac{\partial L}{\partial \dot{x}(t)}    \Big)\delta x(t) +\frac{\partial L}{\partial \dot{x}(t)}\delta x(t_f)  \\
&=\int_{t_i}^{t_f} dt \Big( \frac{\partial L}{\partial x}-\frac{d}{dt}\frac{\partial L}{\partial \dot{x}(t)}    \Big)\delta x(t)+p(t_f)\delta x(t_f)
\end{eqnarray}

If we define $p(t_f)=p_f$ then 

\begin{equation}
p_f=\frac{\partial S[x_c]}{\partial x_f}
\end{equation}

Now the variation of the propagator means 
\begin{equation}
\delta U(x_f,t_f;x_i,t_i)=\delta x(t_f) \frac{\partial }{\partial x_f} U(x_f,t_f;x_i,t_i)
\end{equation}

Therfore 

\begin{equation}
\frac{\partial }{\partial x_f} U(x_f,t_f;x_i,t_i)=\frac{i}{\hbar}\int_{x(t_i)=x_i}^{x(t_f)=x_f}D[x(t)]p(t_f)e^{\frac{i}{\hbar} S[x(t)]}
\end{equation}

The other property that we need to know is the variation of the action with respect to time. As we know 
\begin{equation}
L(x_f,\dot{x}_f)=\frac{d}{dt}S[x(t)]=\frac{\partial S}{\partial t_f}+\frac{\partial S[x(t)]}{\partial x_f}\frac{d x_f}{dt } =\frac{\partial S}{\partial t_f}+p_f\dot{x}_f
\end{equation}

Thus 
\begin{equation}
\frac{\partial S[x_c]}{\partial t_f}=L(x_f,\dot{x}_f)-p_f \dot{x}_f=-H(x_f,p_f)
\end{equation}

Now we can see that 

\begin{equation}
i\hbar \frac{\partial }{\partial t_f} \int_{x(t_i)=x_i}^{x(t_f)=x_f} [Dx(t)] e^{\frac{i}{\hbar} S[x(t)]}=\int_{x(t_i)=x_i}^{x(t_f)=x_f} [Dx(t)]H(x_f,p_f) e^{\frac{i}{\hbar} S[x(t)]}
\end{equation}

Therefore the Propagator is the kernel or The green function of the Schroodinger equation 

\begin{equation}
[i\hbar \frac{\partial }{\partial t_f}-H(x_f,p_f)]U(x_f,t_f;x_i,t_i)=0
\end{equation}

\section{Evolution of Time-Dependent Hamiltonian}

Discution when the systemn is open ....

As we works in the Semi-classica regime we develope our result end evolution with this approximation.
If we consider the Action and change it's variable to 

\begin{equation}
x(\tau):=y(\tau)+\bar{x}(\tau)
\end{equation}

\begin{eqnarray}
\int Ldt &=S(x(\tau))=S(y(\tau)+\bar{x}(\tau)\\ \nonumber
&= S(\bar{x}(\tau))+\frac{\delta S}{\delta x}|_{\bar{x}} y(\tau)+\frac{1}{2}\frac{{\delta}^2 S}{{\delta x}^2}|_{\bar{x}} {y(\tau)}^2
\end{eqnarray}

Then by semi-classical approximation we have 

\begin{equation}
S(\bar{x}+y)\simeq S(\bar{x})+\frac{1}{2}{{\delta}^2 S}y^2
\end{equation}

the most advantage of the Path Integral for our works lie under this property that the path integral is complittly time-dependent formalism and it's not important does the Hamiltonina is time dependent or not and it's the vital property which we need, in the other hand if we works whit canonical fromalism there is no clear conncetion between the time-dependent perturnbation and tiem-independent one. 

\begin{eqnarray}
U(x_f,t_f;x_i,t_i)&& =\int_{x(t_f)}^{x(t_i)}D[x(t)]e^{\frac{i}{\hbar}S[x(t)]} \\ \nonumber 
 && =\int_{x(t_f)}^{x(t_i)}D[y(t)]e^{\frac{i}{\hbar}[S(\bar{x})+\frac{1}{2}{{\delta}^2 S}y^2]}
\end{eqnarray}

Therfore we found the most important lemma that help us to develope the calculation :

\begin{equation}
U(x_f,t_f;x_i,t_i)=e^{\frac{iS(x_f,t_f;x_i,t_i)}{\hbar}}U(0,t_f;0,t_i)
\end{equation}

which the $U(0,t_f;0,t_i)$ is the propagator of the system that hase the hamiltonian $H=H(t)$  for the especail case that the $x_f=x_i=0$ .

by this relation we can obtain the general porpagator that thar the initial and the final positions are arbitrary.

\section{Element of the Hamiltonian }

In principle we can obtian the path integral formalism by accepting the two fundemental property for propagator. the first one, we consider that the propagator has the Markovian behavior.

We consider a bounded operator in Hilbert space, $U(t,t')$, $t \geq t'$ , which describes the evolution from time $t'$ to time $t$ and  satisfies a 
Markov property in time  \citep{2.0}
\begin{equation}\label{a1000}
U(t,t'')U(t'',t')=U(t,t') \ \ \ \ \ \text{for} \ \ \ \ t \geq t'' \geq t 
\end{equation}
Also we consider $U(t',t')=\mathbf{1}$.
moreover, we assume that U(t,t') is differentiable with a continues derivative. We set 
\begin{equation}
\label{a1}
\frac{\partial U(t,t')}{\partial t}|_{t=t'}=\frac{H(t)}{i\hbar}
\end{equation}
here $\hbar$ is real parameter and as we know later it becomes Planck's constant. With this two fundamental properties we can obtain interesting result. By differentiating the Eq.\ref{a1000} with respect to t and take the $t''=t$ we find 
\begin{equation}\label{a2}
i\hbar \frac{\partial U}{\partial t}(t,t')=H(t)U(t,t')
\end{equation}

Now we use from this Identity to estimate the Hamiltonian

\begin{equation}
\langle y |   \frac{\partial U(t,t')}{\partial t}|_{t=t'}  |x \rangle = \frac{1}{i\hbar}\langle y |H(t) |x \rangle 
\end{equation}

as 
\begin{equation}
\langle y |   \frac{\partial U(t,t')}{\partial t}|_{t=t'}  |x \rangle=\frac{\partial }{\partial t}\langle y |    U(t,t')  |x \rangle |_{t=t'}
\end{equation}
Thus 
\begin{eqnarray}
\langle y |H(t) |x \rangle &&=i\hbar \{   [\frac{\partial }{\partial t_f }e^{i\frac{S(x_f,t_f;x_i,t_i)}{\hbar}}U(0,t_f;0,t_i)]|_{t_i=t_f} \\ \nonumber && +[
e^{\frac{S(x_f,t_f;x_i,t_i)}{\hbar}} \frac{\partial }{\partial t_f}U(0,t_f;0,t_i)]|_{t_i=t_f}\}
\end{eqnarray}

Hence  we find that

\begin{eqnarray}
\langle y |H(t) |x \rangle &=i\hbar \{   -\frac{i}{\hbar}H(x_f,p_f) \\ \nonumber & +[
 \frac{\partial }{\partial t_f}U(0,t_f;0,t_i)]|_{t_i=t_f}\}
\end{eqnarray}

here for obtaining the result we need to know the $\frac{\partial }{\partial t_i}U(0,t_f;0,t_i)]$ that by estimating for our system we can easily dervive from it. But as we estimated,  the propagotor for $x_f=x_i=0$  is 

\begin{equation}
U(0,t_f;0,t_i ) =(\frac{\omega}{\pi \hbar})^{\frac{1}{2}} e^{-\frac{i\omega (t_f-t_i)}{2}}e^{-\Gamma (t_f-t_i)}
\end{equation}
Where $\Gamma$ can given by 
\begin{equation}
\Gamma=\hbar |K|e^{-\frac{S_0}{\hbar}}
\end{equation}

Thus 

\begin{equation}
\frac{\partial }{\partial t_f}U(0,t_f;0,t_i)]|_{t_i=t_f}=(\frac{\omega}{\pi \hbar})^{\frac{1}{2}}[-\frac{i\omega}{2}-\Gamma ]
\end{equation}

In our case as the  second derivstion of time dependent potential,$V''$, is independent for time, this estimation is likes to time independent mode.   

\begin{eqnarray}
\langle y |H(t) |x \rangle &=i\hbar \{   -\frac{i}{\hbar}H(x_f,p_f) \\ \nonumber & +[
 (\frac{\omega}{\pi \hbar})^{\frac{1}{2}}[-\frac{i\omega}{2}-\Gamma ]\}
\end{eqnarray}

now if the $|n\rangle $ and $|m\rangle$ be the the two metastable of the system that are time-dependent foundemantaly, then

\begin{eqnarray}
\langle m|H(t) |n \rangle && = \int_x \int_y  dx dy\langle m |x\rangle \langle x|H(t)| y \rangle \langle y|n \rangle \\
&& =\int_x \int_y m^*(x) \langle x|H(t)| y \rangle  n(y)\nonumber 
\end{eqnarray}
 
Or 

\begin{eqnarray}
\langle m|H(t) |n \rangle && = \int_x \int_y  dx dy\langle m |x\rangle \langle x|H(t)| y \rangle \langle y|n \rangle \\   \nonumber 
&& =\int_x \int_y dxdy {\psi}_m^*(x) i\hbar \{   -\frac{i}{\hbar}H(x_f,p_f) \\ \nonumber && +[
 (\frac{\omega}{\pi \hbar})^{\frac{1}{2}}[-\frac{i\omega}{2}-\Gamma ]\} {\psi}_n(y) 
\end{eqnarray}

\begin{appendices}
 
\section{Wave functions}

As we saw in previous section, estimating the element of the Hamiltonian requires the eigenfunction of the eneregy and we need the corresponding wave functions in the reperesentaion of space. Here we want to obtain the approximated from of them and here we use the semi-classical approximation or in the other word we use the WKB approximation and expand wave function by order of $\hbar$ .

\begin{center}
\begin{figure}[ht]\label{YaZahra03}
\includegraphics[height=55mm]{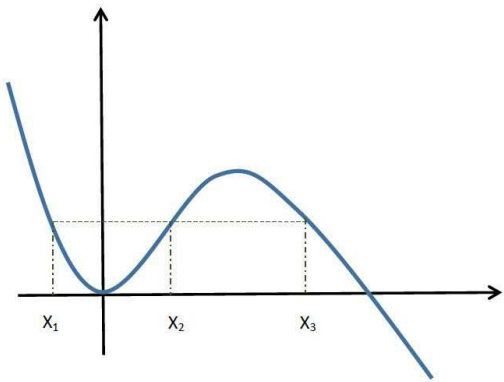}
\caption{ Tilted Washboard potential. }
\end{figure}
\end{center}

In the WKB approximation regime we use the the Pathch function as a uxilary function to connecting the coeficient if the wavefunction in the two side of the returning piont. Therefore near the turning point the wavefunction is near to solution of the diffrential equation which their answer given by airy function 
\begin{equation}
{\psi}_p=a Ai(\alpha x)+ b Bi(\alpha x)
\end{equation}

By defining 
\begin{equation}
\theta:=\frac{1}{\hbar}\int_{x_1}^{x_2} p(x')dx', \ \ \ \ \gamma := \int_{x_2}^{x_3} |p(x)|dx 
\end{equation}

By comparint the coeficient and by using the patch function  near each point we find that

%
%

%
%
%
%
%
%
%
 
 \begin{equation}
\psi(x)\simeq \left\{
\begin{array}{rl}
 \frac{D}{\sqrt{|p(x)|}}e^{-\frac{1}{\hbar}\int_{x}^{x_1}|p(x')|dx'}& x<x_1\\
 -\frac{2D}{\sqrt{p(x)}}\sin[\frac{1}{\hbar}\int_{x}^{x_2}p(x')dx'-\theta -\frac{\pi}{4}]  &   x_1<x<x_2   \\
 \frac{D}{\sqrt{|p(x)|}}[2\cos{\theta}e^{\frac{1}{\hbar}\int_{x_2}^{x}|p(x')|dx'} \\+\sin{\theta} e^{-\frac{1}{\hbar}\int_{x_2}^{x}|p(x')|dx'}]& x_2<x<x_3 \\
\frac{1}{\sqrt{p(x)}}[\frac{D\sin\theta}{e^{\gamma} e^{-i\frac{\pi}{4}}}e^{\frac{i}{\hbar}\int_{x_3}^{x} p(x')dx'}] & x > x_3
\end{array} \right.
\end{equation}

As the amount of enrgy for metastable states has imaginary part, therefore the amplitued of wevefuntion decay gradually and the state dispapear after long time.

 \end{appendices}


\begin{thebibliography}{99}
\bibitem{107}Y. Makhlin, G. Sch¨on and A. Shnirman,\textit{ Rev. Mod. Phys.},
\textbf{73}, 357 ,2001.

\bibitem{108}G. Wendin and V. S. Shumeiko, \textit{Fiz. Nizk. Temp.} 
 [Low Temp. Phys. ],33, 724 ,2007.
\bibitem{109} J.Q. You and F. Nori,\textit{ Phys. Today },text\textbf{58}, 42,2005.
\bibitem{110} J. Clarke and F. K. Wilhelm, \textit{Nature},\textbf{ 453}, 1031, 2008.


\bibitem{106}  M. A. Nielsen and I. L. Chuang,\textit{ Quantum Computation
and Quantum Information} ,Cambridge, 2000.


\bibitem{111}   J. Q.You ,Franco Norim\textit{Superconducting Circuits and Quantum Information}, Physics Today,November, 2005


\bibitem{112}  A.Blais  \textit{Superconducting qubit systems come of age}, Physics today, July, 2009


\bibitem{201}  P. Dirac,\textit{The Lagrangian in Quantum Mechanics}, Physikalische Zeitschrift der Sowjetunion \textbf{3} 64–72,1933

\bibitem{202} Masud Chaichian, Andrei Pavlovich Demichev , \textit{ Path Integrals in Physics}  Volume 1: Stochastic Process and  Quantum Mechanics. Taylor and  Francis. p. 1 ff. ISBN 0-7503-0801-X,2001




\bibitem{1009} [1]S.Coleman. \textit{The Fate of the False Vacuum. 1. Semiclassical Theory} Phys. Rev. D 15, 2929(1977) 

\bibitem{1010} C.G.Callan and S.Coleman,\textit{	The Fate of the False Vacuum. 2. First Quantum Corrections} Phys. Rev. D16, 1762(1977).
\bibitem{1}     C.Coleman, \textit{Aspect of symmetry: selected Erice Lectures of Sidney Coleman},
    Cambridge University Press, Cambridge , 1985.

\bibitem{101} Manin, Yu. I. \textit{Vychislimoe i nevychislimoe [Computable and Noncomputable] (in Russian)}, Sov.Radio. pp. 13–15,1980 

\bibitem{102} Feynman, R. P.\textit{ Simulating physics with computers}, International Journal of Theoretical Physics\textbf{ 21}  467–488,1982 .


\bibitem{103}  Simon, D.R.  ,\textit{ On the power of quantum computation},  Foundations of Computer Science,  1994
\bibitem{104}  Nielsen, Michael A., Chuang, Isaac L. ,\textit{ Quantum Computation and Quantum Information},Cambridge University Press,2001

\bibitem{105} Bennett C.H., Bernstein E., Brassard G., Vazirani U, \textit{The strengths and weaknesses of quantum computation}. SIAM Journal on Computing \textbf{26}: 1510–1523 ,1997.





,



    
\bibitem{2}     Alexander Atland and Ben Simons , second edition \textit{cindensed Matter Field Theory} ,Cambridge University Press,2010 .
\bibitem{2.0}   J.Zinn-Justin, \textit{Path Integrals in Quantum Mechanis },Oxford University Press ,1993.
\bibitem{2.2}   R.Feynman and A.Hibbs, \textit{Quantum Mechanics and Path Integrals} ,McGraw-Hill,New Yourk , 1965.
\bibitem{12}    A. M.Polyakov, \textit{Nucl. Phys} \textbf{B121}, 429 (1997).
\bibitem{13}    J. L. Gervais and B.Sakita, \textit{Phys .Rev .} \textbf{D11}, 2942 (1975).
\bibitem{14}    C. Callan and S.Coleman , \textit{Phys .Rev . } \textbf{D16},1762(1997).
    
\bibitem{15}    J. S. Lsnger, \textit{Ann. Phys.(N.Y)} \textbf{41}, 108 (1967).
\bibitem{16}    L.R. Schulman,  
 \textit{Techniques and Applications of Path Integration} ,Wiley,New York,1981.

\bibitem{18}    A. M. Polyakov,  \textit{Gauge Fields and Strings}, Harwood, 1987.
\bibitem{19}    A. M. Polyakov, \textit{Quark confinement and topology}, \textit{Nucl. Phys.}, \textbf{B120} 429-58(1977).

\bibitem{20}    P. W. Anderson, B. I. Halperin, and C. M. varma, \textit{Anomalous low-temperature thermal properties of glasses and spin glasses,Phil. Mag. } \textbf{25},1-9, 1972

\bibitem{21}    H. Kleinert,  \textit{Path Integrals in Quantum Mechanics, Statistics and  Polymer Physics},World Scientific, Singapore, 1995


\bibitem{22}   C. Grosche and F. Steiner, \textit{Handbook of Feynman path integrals }, Springer,Berlin, Heidelberg, 1998.
\bibitem{23}   L.D. Faddeev, \textit{in Methods in Field Theory}, Les Houches School ,1975.
\bibitem{24}   R.G. Newton, \textit{Scattering theory
Waves and Particles },McGraw-Hill, New York ,1966.
\bibitem{25} G. Wendin and V.S. Shumeiko,\textit{ superconducting  quantum circuits, Qubits and computing wendin review },2005 
\bibitem{30} Hesam Zandi, Shabnam Safaei, Sina Khorasani, Mehdi Fardmanesh, \textit{Study of Junction and Bias Parameters in Readout of Phase Qubits},  Physica C: Superconductivity and its applications , DOI: 10.1016/j.physc.2011.05.002, 2012.
\end{thebibliography}
\end{document}